%% file: ggmet_prl_v4.1.tex
\documentclass[aps,prl,twocolumn,superscriptaddress,floatfix,showpacs]{revtex4}
\usepackage{epsfig}
\usepackage{graphicx}
\usepackage{dcolumn}
\usepackage{amsmath}
\usepackage{xspace}
\graphicspath{./}

\begin{document}
\input{phys_defs.tex}


\title{\bf Search for Supersymmetry with Gauge-Mediated Breaking in Diphoton Events
with Missing Transverse Energy at CDF II}
\preprint{FERMILAB-PUB-09/yyy-E}

\date{\today}

\input{October2009_Authors.tex}

\begin{abstract}
  We present the results of a search for supersymmetry with gauge-mediated  breaking  
and $\NONE\rightarrow\gamma\Gravitino$ in the $\gamma\gamma$+missing transverse energy 
final state. In 2.6$\pm$0.2~\invfb\ of $p{\bar p}$ collisions at $\sqrt{s}$$=$$1.96$~TeV 
recorded by the CDF~II detector we observe no candidate events, consistent with a 
standard model background expectation of 1.4$\pm$0.4 events. We set limits on the cross 
section at the 95\% C.L. and place the world's best limit of 149~\gevc\ on the \none\ mass at
$\tau_{\tilde{\chi}_1^0}$$\ll$1~ns. We also exclude regions in the \none\ mass-lifetime 
plane for $\tau_{\tilde{\chi}_1^0}$$\lesssim$$2$~ns.
\end{abstract}


\maketitle
\newcounter{myctr}

The standard model (SM) of elementary particles has been enormously
successful, but is incomplete.
Theoretical motivations~\cite{gmsb} and the observation of the
`$ee\gamma\gamma+$missing transverse energy ($\mett$)'~\cite{rapdef,eeggmet} candidate event by the CDF experiment during
Run~I at the Fermilab Tevatron provide a compelling rationale to
search for the production and decay of new heavy particles that produce
events with final state photons and \mett\ in collider experiments. Of particular theoretical interest are
supersymmetry (SUSY) models with gauge-mediated SUSY-breaking
(GMSB)~\cite{gmsb}. These models solve the ``naturalness
problem''~\cite{martin} and provide a low-mass dark matter
candidate that is both consistent with inflation and astronomical observations~\cite{astro}.
Since many versions of these models have a similar phenomenology, we consider a
scenario in which 
the lightest neutralino (\neut) decays almost exclusively ($>$96\%) into a photon
($\gamma$) and a weakly
interacting, stable gravitino (\grav). The \grav\ gives rise to \mett\
by leaving the detector without depositing any energy~\cite{snowmass}.
In these models, the \none\ is favored to have a lifetime on the order of a
nanosecond, and the \grav\ is a warm dark matter candidate with a mass in the
range 0.5$<$${m}_{\grav}$$<$1.5~keV$/c^2$~\cite{cosmology}.
Other direct searches~\cite{ggmet,d0search,lep} have constrained the mass of the \neut\ to
be greater than $100$~\gevc\ for various points in parameter space. 
At the Tevatron sparticle production is predicted to result primarily into gaugino
pairs, and the \none\ mass ($m_{\tilde{\chi}_1^0}$) and lifetime
($\tau_{\tilde{\chi}_1^0}$) are the two
most important parameters in determining the final states and their
kinematics~\cite{gmsb}. Different search strategies are
required for \none\ lifetimes above and below about a
nanosecond~\cite{prospects}.

This Letter describes a search for GMSB in which gaugino pairs are
produced and quickly decay
to the $\gamma\gamma$$+$$\mett$$+$$X$ final state, where $X$ denotes other
high-$E_{\rm T}$ final state particles~\cite{eunsin}.
We use a dataset corresponding to an integrated
luminosity of 2.6$\pm$0.2~\invfb\ of $p{\bar p}$ collisions collected with the
CDF~II detector~\cite{CDFII} at $\sqrt{s}$$=$$1.96$~TeV.
This dataset is ten times larger than the one used in our previous
search~\cite{ggmet}. For the first time in this channel we use
a new photon timing system~\cite{nim} and a new model of the \mett\ resolution
({\sc metmodel})~\cite{ggXPRD}. These additions significantly improve our rejection
of backgrounds from instrumental and non-collision sources, which allows us
to considerably enhance the sensitivity of the search for large \none\ masses compared to
other Tevatron searches~\cite{d0search}. We also extend the search by
addressing \none\ lifetimes up to
2~ns, which are favored for larger $m_{\tilde{\chi}_1^0}$.


Here we briefly describe
the aspects of the detector~\cite{CDFII} relevant to this analysis. The magnetic
spectrometer consists of tracking devices that measures the $z$ position and
time of the $\pbarp$ interaction, and the momenta of charged particles inside a
superconducting solenoid magnet. 
The calorimeter consists of electromagnetic (EM) and
hadronic (HAD) compartments and is divided into a central part that surrounds
the solenoid coil
($|\eta|$$<$$1.1$)~\cite{rapdef} and a pair
of end-plugs that cover the region $1.1$$<$$|\eta|$$<$$3.6$.
The calorimeters are used to identify and measure the 4-momenta of photons, electrons, and jets ($j$)~\cite{jets}
and to provide $\mett$ information.
The EM calorimeter is instrumented with a timing system (EMTiming)~\cite{nim} 
that measures the arrival time of photons.

Our analysis begins with diphoton events passing the CDF three-level
trigger. The combined trigger selection efficiency is
effectively 100\%  if both photons have
$|\eta|$$<$$1.1$ and $E_{\rm T}$$>$$13$~GeV~\cite{eunsin,ggXPRD}.
Offline, both photons are required to be in the fiducial part of the calorimeter and 
to pass the standard CDF photon identification and isolation
requirements~\cite{ggmet},
with two minor modifications to remove instrumental and electron
backgrounds~\cite{ggXPRD,delayedPRLD}.
The remaining events are dominated by SM production of $\gamma\gamma$, $\gamma j$
with $j$$\to$$\gamma_{fake}$, and $jj$$\to$$\gamma_{fake}\gamma_{fake}$, where
$\gamma_{fake}$ is any object misidentified as a photon.
To minimize the number of these events with large \mett\ due to calorimeter energy
mismeasurements, we remove events 
where the azimuthal angle between the \mett\ and the second-highest $E_{\rm T}$
photon is $|\Delta\phi|$$<$$0.3$ or if any jet points to an uninstrumented
region of the calorimeter~\cite{ggXPRD}.
We require a primary collision vertex position with $|z_{\rm vertex}|$$<$60~cm 
in order to reduce non-collision backgrounds and to maintain the projective nature of the
photon reconstruction in the calorimeter.
For events with multiple reconstructed vertices we recalculate the $E_{\rm T}$ of
both photons
and \mett\ values if picking a different vertex for them reduces the event
\mett.

Non-collision backgrounds coming from cosmic rays and
beam-related effects can produce  $\gamma\gamma$$+$$\mett$ candidates, and are removed
from the inclusive $\gamma\gamma$  sample using a number of techniques. Photon candidates
from cosmic rays are not correlated in time with collisions. Therefore, events are
removed  
 if the timing of either photon, corrected for average path length
($t_{\gamma}$), indicates a non-collision source~\cite{ggXPRD,delayedPRLD}.
 Photon candidates can also be produced
by beam-related muons that originate upstream of the detector (from the more
intense $p$ beam).
These are suppressed using standard beam halo identification
requirements~\cite{delayedPRLD}. 
A total of 38,053 inclusive $\gamma\gamma$ candidate events pass all the 
selection requirements.

Backgrounds to the $\gamma\gamma$$+$$\mett$ final state from SM $\gamma\gamma$/
$\gamma\gamma_{fake}$/$\gamma_{fake}\gamma_{fake}$ and fake $\mett$ arise due to
energy mismeasurements in the calorimeter or to event reconstruction pathologies.
We use the {\sc metmodel}~\cite{ggXPRD} to select events with real and significant $\mett$, as part of the optimization, and to predict the contribution 
of SM backgrounds with fake $\mett$ due to normal energy measurement fluctuations.
This algorithm considers the clustered (jets) and unclustered energy in
the event and calculates the probability for fluctuations in the energy measurement to produce
${E\!\!\!\!/_{T}~\!\!\!\!^{fluct}}$ equivalent to or larger than the measured
$\mett$, $P_{{E\!\!\!\!/_{T}~\!\!\!\!^{fluct}}\geq\mett}$. This
probability is then used to define a $\mett$-significance as $-$$\log _{10}
\left(P_{{E\!\!\!\!/_{T}~\!\!\!\!^{fluct}}\geq\mett}\right)$. Events with true and fake $\mett$ of the
same value have, on average, different $\mett$-significance.
We use pseudo-experiments to estimate the expected $\mett$-significance distribution for SM events with fake
\mett, and the number of
mismeasured events above a given $\mett$-significance requirement.
The jets and unclustered energy are smeared according to
their resolution functions in the event.
The systematic uncertainty in the {\sc metmodel} is dominated by the uncertainty
in the resolution functions.

The {\sc metmodel} does not account for reconstruction pathologies in SM events
without intrinsic \mett, such as a
wrong choice of the primary interaction vertex or tri-photon events with a lost
photon.
To obtain the prediction for this background we model SM kinematics and event
reconstruction using a $\gamma\gamma$ sample generated with a {\sc pythia} Monte Carlo (MC)~\cite{pythia}
that incorporates a detector simulation~\cite{geant}. 
Since the pathologies from $\gamma j$ and $jj$ sources are similar in nature,
but not included directly in the simulation, we normalize the sample to the number
of events in the inclusive $\gamma\gamma$ data sample.
We subtract the expectations for energy mismeasurement fluctuations in
the MC to avoid double counting.
Uncertainties are dominated by the statistics of the MC sample, but also include
the small differences between the measured response of the {\sc metmodel} to MC
simulation events and real data.

Electroweak production of $W$ and $Z$ bosons which decay to leptons can also 
produce the
$\gamma\gamma$$+$$\mett$ signature where one or more of the photons can be fake, but
the \mett\ is due to one or more neutrinos. To estimate the contribution from
these backgrounds we use MC simulations
normalized to their theoretical cross sections, taking into account all the leptonic
decay modes. 
The Baur MC~\cite{baur} is used to simulate $W\gamma$ and $Z\gamma$ production
and decay where initial and final state radiation (ISR/FSR) produce 
$W/Z$$+$$\gamma\gamma$ events. The {\sc pythia} MC is used to simulate
backgrounds where both photons are fakes: namely, $W$ and $Z$, with photons from ISR/FSR
removed, and $t\bar{t}$ sources. 
To minimize the dependence of our predictions on potential ``MC-data'' differences
we scale our MC predictions to the observed number of $e\gamma$
events~\cite{ggXPRD} in data
where we use the same diphoton triggers and analysis selection procedures
used to select the inclusive $\gamma\gamma$ sample.
Uncertainties are dominated by the statistics of the MC and $e\gamma$
normalization data sample.

\begin{figure}[tbp]
\includegraphics[width=1.\linewidth]{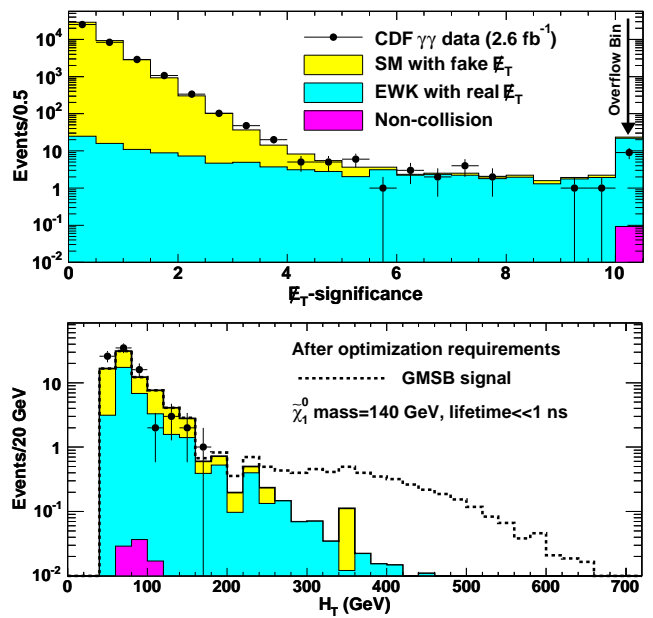}
  \caption{The top plot shows the $\mett$-significance distribution for the
inclusive $\gamma\gamma$ candidate sample, along with the background predictions.
The bottom plot shows the predicted $H_{\rm T}$ distribution after all but the
final $H_{\rm T}$ requirement.}
  \label{fig:prl1}
\end{figure}

Non-collision backgrounds are estimated using the data.
We identify a cosmic-enhanced sample by using the selected inclusive $\gamma\gamma$ 
sample, but requiring one of the photons to
have $t_{\gamma}$$>$$25$~ns. 
Similarly, we create a beam halo-enhanced sample from events that were filtered
out from our signal sample by the beam halo rejection requirements~\cite{delayedPRLD}. We
estimate the non-collision background events in the signal region using 
extrapolation techniques and the measured efficiencies of the non-collision
rejection requirements~\cite{ggXPRD}.
The uncertainties on both non-collision background estimates are dominated by
the statistical
uncertainty on the number of identified events. Figure~\ref{fig:prl1} (top)
shows the $\mett$-significance distribution for the inclusive $\gamma\gamma$
sample, along with the predictions
for all the backgrounds.

We estimate the sensitivity to heavy, neutral particles that decay
to photons using the GMSB reference model~\cite{snowmass}
in the mass-lifetime range, $75$$\leq$$m_{\tilde{\chi}_1^0}$$\leq$$150$~GeV and
$\tau_{\tilde{\chi}_1^0}$$\lesssim$$2$~ns. 
Events from all SUSY processes considered~\cite{simeon} are simulated with
{\sc pythia} followed by a detector simulation. 
The fraction of \neut\ decays that
occur in the detector volume, and thus the acceptance, depend on both the
lifetime and the masses of the sparticles~\cite{prospects}.
The total systematic uncertainty on the acceptance, after all kinematic
requirements (discussed below), is estimated to
be 7\%, dominated by the uncertainty in the photon identification efficiency (2.5\% per photon). Other
significant contributions come from uncertainties on ISR/FSR (4\%), jet energy measurement
(2\%), 
$\mett$-significance parameterizations (1\%) and parton distribution functions
(PDFs, 1\%).

We determine the final kinematic selection requirements by optimizing the mean expected 95\%
confidence level (C.L.) cross
section limit using a no-signal assumption, before looking at the data in the
signal region~\cite{Boos}.  To
compute the predicted cross section upper
limit we combine the luminosity, the acceptance, and the background
estimates with their systematic uncertainties using a Bayesian
method~\cite{junk}. The predicted limits are optimized by
simultaneously varying the selection requirements for $\mett$-significance,
$H_{\rm T}$ (scalar sum of  $E_{\rm T}$
of photons, jets, and $\mett$), and the azimuthal angle between the two leading photons, 
$\Delta\phi(\gamma_{1}, \gamma_{2})$. The large $\mett$-significance requirement
eliminates most of the SM background with fake
$\mett$.  GMSB production is dominated by heavy gaugino pairs which decay to
high-$E_{\rm T}$ light final state particles via cascade decays. The GMSB signal has, on
average, larger $H_{\rm T}$ compared
to SM backgrounds so that an $H_{\rm T}$ requirement can remove these backgrounds effectively.
Electroweak backgrounds with large $H_{\rm T}$ typically consist of a high-$E_{\rm T}$ photon recoiling against
$W$$\to$$e\nu$, identified as $\gamma_{fake}\mett$, which means the gauge boson decay is highly boosted. 
Thus, the two photon candidates in the final state are mostly back-to-back.
The SM backgrounds with fake \mett\ and large $H_T$ also have photons which are
mostly back-to-back; the $\Delta\phi(\gamma_{1}, \gamma_{2})$ requirement, therefore, reduces both these backgrounds.

The optimal set of requirements is slightly different for each point in the
$\tau_{\tilde{\chi}_1^0}$ vs. $m_{\tilde{\chi}_1^0}$ space
considered.
We choose a single set of requirements to
maximize the region where the predicted production cross
section at next-to-leading order~\cite{kfactors} is above the expected 95\% C.L. cross
section limit. The exclusion region also takes into account the production cross
section uncertainties, which are dominated by the PDFs (7\%) and the
renormalization scale (3\%).
We find the optimal set of requirements, before unblinding the signal region, to be:
$\mett$-significance$>$$3$, $H_{\rm T}$$>$$200$~GeV, and
$\Delta\phi(\gamma_{1},\gamma_{2})$$<$$\pi$$-$$0.35$.
With these requirements we predict 1.4$\pm$0.4 background events, 0.9$\pm$0.4 of
which are from electroweak sources (dominated by $Z\gamma\gamma$
production) with real $\mett$, 0.5$\pm$0.2 from SM with fake $\mett$, and 
0.001$^{+0.008}_{-0.001}$ from non-collision sources. The acceptance for
$m_{\tilde{\chi}_1^0}$$=$$140$~\gevc\ and
$\tau_{\tilde{\chi}_1^0}$$\ll$$1$~ns is estimated to be 7.8$\pm$0.6\%.

\begin{figure}[tbp]
    \includegraphics[width=1.\linewidth]{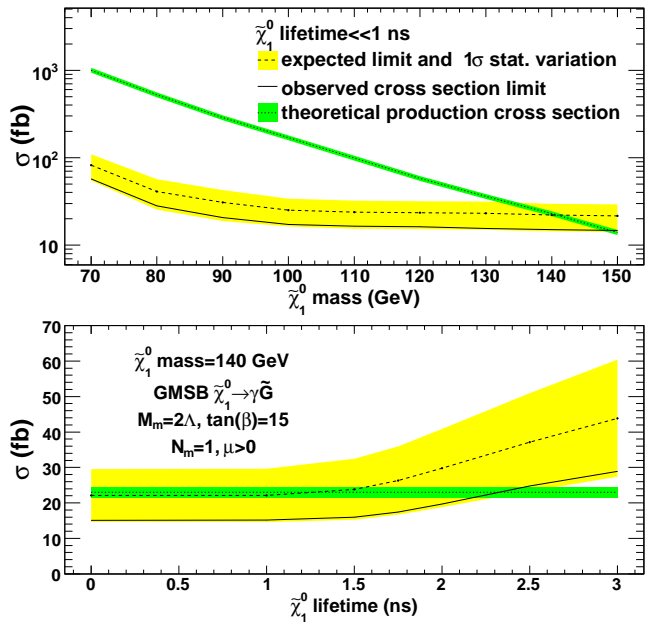}
  \caption{The predicted and observed 95\% C.L. cross section upper limits as
    a function of the \none\ mass at $\tau_{\tilde{\chi}_1^0}$$\ll$$1$~ns (top) and as a
    function of the \none\ lifetime at $m_{\tilde{\chi}_1^0}$$=$$140$~GeV/$c^{2}$ (bottom).
Indicated
    in green (darker shading) is the production cross section, along with its 8.0\%
uncertainty-band.
    In yellow (lighter shading) is the RMS variation on the expected cross section
    limit.}
  \label{fig:prl2}
\end{figure}

\begin{figure}[bt]
    \includegraphics[width=1\linewidth]{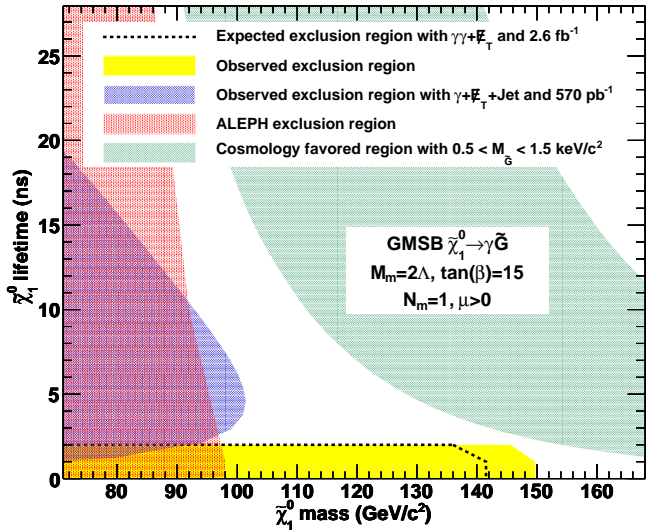}
  \caption{The predicted and observed exclusion region along with the limits
found in~\cite{lep,delayedPRLD}. 
The shaded band shows the parameter space where
$0.5$$<$$m_{\Gravitino}$$<$$1.5~{\rm keV}/c^{2}$, favored by cosmological
models~\cite{cosmology}.}
  \label{fig:prl3}
\end{figure}

No events in the data pass the final event selection. 
The predicted $H_{\rm T}$ distribution is shown in Fig.~\ref{fig:prl1}
(bottom), after all but the final $H_{\rm T}$ requirement. The
data are consistent with the no-signal hypothesis and are well modeled by SM backgrounds
alone. We set cross section limits as a function of
$m_{\tilde{\chi}_1^0}$ and $\tau_{\tilde{\chi}_1^0}$, respectively, as shown in
Fig.~\ref{fig:prl2}. 
The $m_{\tilde{\chi}_1^0}$ reach, based
on the predicted and observed number of events for
$\tau_{\tilde{\chi}_1^0}$$\ll$$1$~ns, 
is 141 \gevc\ and 149 \gevc\ respectively.
These limits significantly extend the search sensitivity beyond the results of
D0~\cite{d0search}, expand the results to include exclusions for
$\tau_{\tilde{\chi}_1^0}$$\leq$$2$~ns, and, when combined with the complementary
limits from CDF
and LEP~\cite{delayedPRLD,lep}, cover the region shown in Fig.~\ref{fig:prl3}.

In conclusion, we have performed an optimized search for heavy,
neutral particles that decay to photons in the $\gamma\gamma$$+$$\mett$ final state using
2.6$\pm$0.2~fb$^{-1}$ of data. There is no excess of events beyond expectations.
We set cross section limits using a GMSB
model with \NTGTGR, and find an exclusion region in the
$\tau_{\tilde{\chi}_1^0}$-$m_{\tilde{\chi}_1^0}$ 
plane with the world's best 95\% C.L. lower limit on
the \none\ mass of 149~\gevc\ at
$\tau_{\tilde{\chi}_1^0}$$\ll$$1$~ns. 
By the end of Run II, with an integrated luminosity of 10~\invfb, we estimate a
mass reach of $\simeq 160$~\gevc.
\begin{acknowledgments}
  We thank the Fermilab staff and the technical staffs of the
  participating institutions for their vital contributions. This work
  was supported by the U.S. Department of Energy and National Science
  Foundation; the Italian Istituto Nazionale di Fisica Nucleare; the
  Ministry of Education, Culture, Sports, Science and Technology of
  Japan; the Natural Sciences and Engineering Research Council of
  Canada; the National Science Council of the Republic of China; the
  Swiss National Science Foundation; the A.P. Sloan Foundation; the
  Bundesministerium f\"ur Bildung und Forschung, Germany; the World Class
  University Program, the National Research Foundation of Korea;
  the Science and Technology Facilities Council and the Royal Society, UK;
  the Institut National de Physique Nucleaire et Physique des Particules/CNRS;
  the Russian Foundation for Basic Research; the Ministerio de Ciencia e
  Innovaci\'{o}n, and Programa Consolider-Ingenio 2010, Spain; the Slovak R\&D
  Agency; and the Academy of Finland. 
\end{acknowledgments}

\end{document}

%% file: phys_defs.tex
\newcommand{\invfb}{$\rm fb^{-1}$} 

\newcommand{\photon}{$\gamma$}
\newcommand{\ett}{\Et}
\newcommand{\ptt}{\mbox{$P_T$}}
\newcommand{\PT}{\ptt}
\newcommand{\NONE}{\mbox{$\widetilde{\chi}_1^0$}}
\newcommand{\NTWO}{\mbox{$\widetilde{\chi}_2^0$}}
\newcommand{\CONE}{\mbox{$\widetilde{\chi}_1^{\pm}$}}
\newcommand{\CONEP}{\mbox{$\widetilde{\chi}_1^{+}$}}
\newcommand{\CONEM}{\mbox{$\widetilde{\chi}_1^{-}$}}
\newcommand{\none}{\NONE}
\newcommand{\ntwo}{\NTWO}
\newcommand{\cone}{\CONE}
\newcommand{\coneb}{\CONEB}
\newcommand{\NTGNO}{\mbox{$\NTWO \rightarrow \gamma \NONE$}}
\newcommand{\NTGTGR}{\mbox{$\NONE \rightarrow \gamma \Gravitino$}}
\newcommand{\tanbeta}{\tan\beta}

\newcommand{\gevc}{GeV/c$^2$}

\newcommand{\mysect}[1]{\addtocounter{myctr}{1} \noindent \underline{Paragraph {\arabic{myctr}}: #1} \\}
\newcommand{\neut}{$ \tilde{\chi}_1^0 $}
\newcommand{\grav}{\ensuremath{\tilde{G}}}
\newcommand{\mett}{\mbox{$E\!\!\!\!/_{\rm T}$}}
\newcommand{\met}{\mett}
\newcommand{\eeggmett}{$ee\gamma\gamma\mett $}
\newcommand{\ggmett}{$\gamma\gamma\mett $}
\newcommand{\gmett}{$\gamma$\met}
\newcommand{\scinotn}[2]{\mbox{${#1}\times 10^{#2}$}}
\newcommand{\sumet}{$\sum E_{\rm T} $}
\newcommand{\sumpt}{$\sum P_{\rm T} $}
\newcommand{\dphicosm}{$\Delta\phi $}
\newcommand{\chisq}{${\chi}^2 $}
\newcommand{\etal}{{\it et al.}}

\newcommand{\zpos}{ {\it Z} }
\newcommand{\trkz}{$Z_{\rm 0}$}
\newcommand{\trkt}{$T_{\rm 0}$}

\def\roots{{\sqrt s}}
\def \Et {$E_{\rm T}$}
\def \Etg {E_{\rm T}^{\gamma}}
\def\Z{{ Z^0}}
\def\selectron{\widetilde{e}}
\def\goes{\rightarrow}
\def\Gravitino{\widetilde{G}}
\def\pbarp{p{\bar p}}
\def\degrees{^\circ}
\def\deg{^\circ}
\def\ppbar{p{\bar p}}

\def\W{{ $W^{\pm}$}}
\def\Wele{ $W \rightarrow e\nu$}
\def\Zee{ $Z \rightarrow e^+e^-$}

\def\Journal#1#2#3#4{{#1} {\bf #2} (#4) #3}
\def\PrePrint#1{\mbox{hep-ph/#1}}
\def\PRL{\rm Phys. Rev. Lett.}
\def\PRD{{\rm Phys. Rev.} D}
\def\NIM#1#2#3{\rm Nucl. Instr. and Meth A{#1}~(#2)~#3}

\def\rphi{r-$\phi$}                                                                           
\def\rz{r-z}
\def\and{\\ \vspace{0.3cm}}

%% file: October2009_Authors.tex
\affiliation{Institute of Physics, Academia Sinica, Taipei, Taiwan 11529, Republic of China} 
\affiliation{Argonne National Laboratory, Argonne, Illinois 60439} 
\affiliation{University of Athens, 157 71 Athens, Greece} 
\affiliation{Institut de Fisica d'Altes Energies, Universitat Autonoma de Barcelona, E-08193, Bellaterra (Barcelona), Spain} 
\affiliation{Baylor University, Waco, Texas  76798} 
\affiliation{Istituto Nazionale di Fisica Nucleare Bologna, $^{cc}$University of Bologna, I-40127 Bologna, Italy} 
\affiliation{Brandeis University, Waltham, Massachusetts 02254} 
\affiliation{University of California, Davis, Davis, California  95616} 
\affiliation{University of California, Los Angeles, Los Angeles, California  90024} 
\affiliation{University of California, San Diego, La Jolla, California  92093} 
\affiliation{University of California, Santa Barbara, Santa Barbara, California 93106} 
\affiliation{Instituto de Fisica de Cantabria, CSIC-University of Cantabria, 39005 Santander, Spain} 
\affiliation{Carnegie Mellon University, Pittsburgh, PA  15213} 
\affiliation{Enrico Fermi Institute, University of Chicago, Chicago, Illinois 60637}
\affiliation{Comenius University, 842 48 Bratislava, Slovakia; Institute of Experimental Physics, 040 01 Kosice, Slovakia} 
\affiliation{Joint Institute for Nuclear Research, RU-141980 Dubna, Russia} 
\affiliation{Duke University, Durham, North Carolina  27708} 
\affiliation{Fermi National Accelerator Laboratory, Batavia, Illinois 60510} 
\affiliation{University of Florida, Gainesville, Florida  32611} 
\affiliation{Laboratori Nazionali di Frascati, Istituto Nazionale di Fisica Nucleare, I-00044 Frascati, Italy} 
\affiliation{University of Geneva, CH-1211 Geneva 4, Switzerland} 
\affiliation{Glasgow University, Glasgow G12 8QQ, United Kingdom} 
\affiliation{Harvard University, Cambridge, Massachusetts 02138} 
\affiliation{Division of High Energy Physics, Department of Physics, University of Helsinki and Helsinki Institute of Physics, FIN-00014, Helsinki, Finland} 
\affiliation{University of Illinois, Urbana, Illinois 61801} 
\affiliation{The Johns Hopkins University, Baltimore, Maryland 21218} 
\affiliation{Institut f\"{u}r Experimentelle Kernphysik, Karlsruhe Institute of Technology, D-76131 Karlsruhe, Germany} 
\affiliation{Center for High Energy Physics: Kyungpook National University, Daegu 702-701, Korea; Seoul National University, Seoul 151-742, Korea; Sungkyunkwan University, Suwon 440-746, Korea; Korea Institute of Science and Technology Information, Daejeon 305-806, Korea; Chonnam National University, Gwangju 500-757, Korea; Chonbuk National University, Jeonju 561-756, Korea} 
\affiliation{Ernest Orlando Lawrence Berkeley National Laboratory, Berkeley, California 94720} 
\affiliation{University of Liverpool, Liverpool L69 7ZE, United Kingdom} 
\affiliation{University College London, London WC1E 6BT, United Kingdom} 
\affiliation{Centro de Investigaciones Energeticas Medioambientales y Tecnologicas, E-28040 Madrid, Spain} 
\affiliation{Massachusetts Institute of Technology, Cambridge, Massachusetts  02139} 
\affiliation{Institute of Particle Physics: McGill University, Montr\'{e}al, Qu\'{e}bec, Canada H3A~2T8; Simon Fraser University, Burnaby, British Columbia, Canada V5A~1S6; University of Toronto, Toronto, Ontario, Canada M5S~1A7; and TRIUMF, Vancouver, British Columbia, Canada V6T~2A3} 
\affiliation{University of Michigan, Ann Arbor, Michigan 48109} 
\affiliation{Michigan State University, East Lansing, Michigan  48824}
\affiliation{Institution for Theoretical and Experimental Physics, ITEP, Moscow 117259, Russia} 
\affiliation{University of New Mexico, Albuquerque, New Mexico 87131} 
\affiliation{Northwestern University, Evanston, Illinois  60208} 
\affiliation{The Ohio State University, Columbus, Ohio  43210} 
\affiliation{Okayama University, Okayama 700-8530, Japan} 
\affiliation{Osaka City University, Osaka 588, Japan} 
\affiliation{University of Oxford, Oxford OX1 3RH, United Kingdom} 
\affiliation{Istituto Nazionale di Fisica Nucleare, Sezione di Padova-Trento, $^{dd}$University of Padova, I-35131 Padova, Italy} 
\affiliation{LPNHE, Universite Pierre et Marie Curie/IN2P3-CNRS, UMR7585, Paris, F-75252 France} 
\affiliation{University of Pennsylvania, Philadelphia, Pennsylvania 19104}
\affiliation{Istituto Nazionale di Fisica Nucleare Pisa, $^{ee}$University of Pisa, $^{ff}$University of Siena and $^{gg}$Scuola Normale Superiore, I-56127 Pisa, Italy} 
\affiliation{University of Pittsburgh, Pittsburgh, Pennsylvania 15260} 
\affiliation{Purdue University, West Lafayette, Indiana 47907} 
\affiliation{University of Rochester, Rochester, New York 14627} 
\affiliation{The Rockefeller University, New York, New York 10021} 
\affiliation{Istituto Nazionale di Fisica Nucleare, Sezione di Roma 1, $^{hh}$Sapienza Universit\`{a} di Roma, I-00185 Roma, Italy} 

\affiliation{Rutgers University, Piscataway, New Jersey 08855} 
\affiliation{Texas A\&M University, College Station, Texas 77843} 
\affiliation{Istituto Nazionale di Fisica Nucleare Trieste/Udine, I-34100 Trieste, $^{ii}$University of Trieste/Udine, I-33100 Udine, Italy} 
\affiliation{University of Tsukuba, Tsukuba, Ibaraki 305, Japan} 
\affiliation{Tufts University, Medford, Massachusetts 02155} 
\affiliation{Waseda University, Tokyo 169, Japan} 
\affiliation{Wayne State University, Detroit, Michigan  48201} 
\affiliation{University of Wisconsin, Madison, Wisconsin 53706} 
\affiliation{Yale University, New Haven, Connecticut 06520} 
\author{T.~Aaltonen}
\affiliation{Division of High Energy Physics, Department of Physics, University of Helsinki and Helsinki Institute of Physics, FIN-00014, Helsinki, Finland}
\author{J.~Adelman}
\affiliation{Enrico Fermi Institute, University of Chicago, Chicago, Illinois 60637}
\author{B.~\'{A}lvarez~Gonz\'{a}lez$^v$}
\affiliation{Instituto de Fisica de Cantabria, CSIC-University of Cantabria, 39005 Santander, Spain}
\author{S.~Amerio$^{dd}$}
\affiliation{Istituto Nazionale di Fisica Nucleare, Sezione di Padova-Trento, $^{dd}$University of Padova, I-35131 Padova, Italy} 

\author{D.~Amidei}
\affiliation{University of Michigan, Ann Arbor, Michigan 48109}
\author{A.~Anastassov}
\affiliation{Northwestern University, Evanston, Illinois  60208}
\author{A.~Annovi}
\affiliation{Laboratori Nazionali di Frascati, Istituto Nazionale di Fisica Nucleare, I-00044 Frascati, Italy}
\author{J.~Antos}
\affiliation{Comenius University, 842 48 Bratislava, Slovakia; Institute of Experimental Physics, 040 01 Kosice, Slovakia}
\author{G.~Apollinari}
\affiliation{Fermi National Accelerator Laboratory, Batavia, Illinois 60510}
\author{A.~Apresyan}
\affiliation{Purdue University, West Lafayette, Indiana 47907}
% ...

\collaboration{CDF Collaboration\footnote{With visitors from $^a$University of Massachusetts Amherst, Amherst, Massachusetts 01003,
$^b$Universiteit Antwerpen, B-2610 Antwerp, Belgium, 
$^c$University of Bristol, Bristol BS8 1TL, United Kingdom,
$^d$Chinese Academy of Sciences, Beijing 100864, China, 
$^e$Istituto Nazionale di Fisica Nucleare, Sezione di Cagliari, 09042 Monserrato (Cagliari), Italy,
$^f$University of California Irvine, Irvine, CA  92697, 
$^g$University of California Santa Cruz, Santa Cruz, CA  95064, 
$^h$Cornell University, Ithaca, NY  14853, 
$^i$University of Cyprus, Nicosia CY-1678, Cyprus, 
$^j$University College Dublin, Dublin 4, Ireland,
$^k$University of Edinburgh, Edinburgh EH9 3JZ, United Kingdom, 
$^l$University of Fukui, Fukui City, Fukui Prefecture, Japan 910-0017
$^m$Kinki University, Higashi-Osaka City, Japan 577-8502
$^n$Universidad Iberoamericana, Mexico D.F., Mexico,
$^o$University of Iowa, Iowa City, IA  52242,
$^p$Kansas State University, Manhattan, KS 66506
$^q$Queen Mary, University of London, London, E1 4NS, England,
$^r$University of Manchester, Manchester M13 9PL, England,
$^s$Muons, Inc., Batavia, IL 60510, 
$^t$Nagasaki Institute of Applied Science, Nagasaki, Japan, 
$^u$University of Notre Dame, Notre Dame, IN 46556,
$^v$University de Oviedo, E-33007 Oviedo, Spain, 
$^w$Texas Tech University, Lubbock, TX  79609, 
$^x$IFIC(CSIC-Universitat de Valencia), 56071 Valencia, Spain,
$^y$Universidad Tecnica Federico Santa Maria, 110v Valparaiso, Chile,
$^z$University of Virginia, Charlottesville, VA  22906
$^{aa}$Bergische Universit\"at Wuppertal, 42097 Wuppertal, Germany,
$^{bb}$Yarmouk University, Irbid 211-63, Jordan
$^{jj}$On leave from J.~Stefan Institute, Ljubljana, Slovenia, 
}}
\noaffiliation

%% file: ggmet_prl_v4.1.bbl
\begin{thebibliography}{99}
\bibitem{rapdef} We use a cylindrical coordinate system in which the
  proton beam travels along the $z$-axis, $\theta$ is the polar angle,
  $\phi$ is the azimuthal angle relative to the horizontal plane, and $\eta$$=$$-\ln \tan(\theta/2)$. The
  transverse energy and momentum are defined as $E_T$$=$$E\sin\theta$ and
  $p_T$$=$$p \sin \theta$ where $E$ is the energy measured by the
  calorimeter and $p$ the momentum measured in the tracking system.
  $\mett$$=$$|-\sum_i E_T^i \vec{n_i}|$ where $\vec{n_i}$ is a unit vector
  that points from the interaction vertex to the $i^{\rm th}$ calorimeter
  tower in the transverse plane.
\end{thebibliography}
